\documentclass[aps,pra,epsfig,graphics,showpacs,twocolumn,floatfix,mathbbm,a4paper]{revtex4}
\newcommand{\id}{{\sf 1 \hspace{-0.3ex} \rule{0.1ex}{1.52ex}\rule[-.01ex]{0.3ex}{0.1ex}}}
\newcommand{\idsmall}{{\sf 1 \hspace{-0.2ex} \rule{0.1ex}{1ex}\rule[-.01ex]{0.25ex}{0.1ex}}}
\usepackage{amsmath,amsfonts,amssymb,graphics,graphicx,epsfig,color,times,bbm}

\begin{document}
\bibliographystyle{apsrev}
\date{\today}
\title{Local copying of orthogonal entangled quantum states}
\author{Fabio Anselmi}
\affiliation{School of Physics, Astronomy and Mathematics,
University of Hertfordshire,
       Hatfield AL10 9AB, Hertfordshire, UK}
\author{Anthony Chefles\footnote{Present address: Department of Mathematical Physics, NUI Maynooth, Maynooth, Co. Kildare, Ireland}}
\affiliation{School of Physics, Astronomy and Mathematics,
University of Hertfordshire,
       Hatfield AL10 9AB, Hertfordshire, UK}
\author{Martin B. Plenio}
\affiliation{QOLS, The Blackett Laboratory, Imperial College
London, \it Prince Consort Rd., London SW7 2BW, UK}
\begin{abstract}
\vspace{0.5cm}
 In classical information theory one can, in
principle, produce a perfect copy of any input state. In quantum
information theory, the no cloning theorem prohibits exact copying
of nonorthogonal states. Moreover, if we wish to copy
multiparticle entangled states and can perform only local
operations and classical communication (LOCC), then further
restrictions apply. We investigate the problem of copying
orthogonal, entangled quantum states with an entangled blank state
under the restriction to LOCC.  Throughout, the subsystems have
finite dimension $D$. We show that if all of the states to be
copied are non-maximally entangled, then novel LOCC copying
procedures based on entanglement catalysis are possible. We then
study in detail the LOCC copying problem where both the blank
state and at least one of the states to be copied are maximally
entangled.  For this to be possible, we find that all the states
to be copied must be maximally entangled. We obtain a necessary
and sufficient condition for LOCC copying under these conditions.
For two orthogonal, maximally entangled states, we provide the
general solution to this condition. We use it to show that for
$D=2,3$, any pair of orthogonal, maximally entangled states can be
locally copied using a maximally entangled blank state. However,
we also show that for any $D$ which is not prime, one can
construct pairs of such states for which this is impossible.
\vspace{0.5cm}

\end{abstract}
 \maketitle
\newtheorem{postulate}{Postulate}
\newtheorem{theorem}{Theorem}
\newtheorem{definition}[theorem]{Definition}
\newtheorem{lemma}[theorem]{Lemma}
\newtheorem{example}[theorem]{Example}

\section{Introduction}
\label{sec1} The no cloning theorem of Wootters and Zurek
\cite{Wootters} and Dieks \cite{Dieks} prohibits the creation of
perfect copies of nonorthogonal quantum states.  This famous
result has profound implications for quantum communications, e.g.
the security of quantum cryptography \cite{Cr}. It is also well
known that any set of orthogonal states can be perfectly copied in
principle. However it is not known how well this can be achieved
if there are restrictions on the set of possible quantum
operations.

A common scenario in quantum information processing and
communications is where a multiparticle, possibly entangled state
is distributed among a number of spatially separated parties. Each
of these parties can perform arbitrary local operations on the
subsystems they possess. However, they can only send classical
information to each other. When this is the case, the parties are
restricted to performing local (quantum) operations and classical
communication (LOCC). There has  been a considerable amount of
activity devoted to understanding the properties of LOCC
operations. Certain specific quantum information processing tasks,
such as entanglement distillation and, more recently, state
discrimination, have been the focus of a particularly large amount
of attention with respect to the LOCC constraint. In this paper,
we investigate the problem of copying orthogonal, entangled,
quantum states under these conditions.

Quantum copying and quantum state discrimination are closely
related operations \cite{Review}.  In the study of state
discrimination under the LOCC constraint, it has been found that
any pair of orthogonal, entangled, pure, bipartite states can be
perfectly discriminated by LOCC \cite{Walgate1}.  This is not
generally possible for more than two states. Also, it has been
found that any two nonorthogonal, entangled, pure, bipartite
states can be optimally discriminated by LOCC
\cite{Virmani,Chen,Ji}. Again, this is not generally possible for
more than two states \cite{USDLOCC}.

We see that the LOCC constraint imposes restrictions on the number
of states for which certain discrimination tasks are possible.
Given that copying is closely related to state discrimination, we
might imagine that the LOCC constraint could also affect the
number of states for which certain copying procedures are
possible. We will show in this paper that this is indeed the case.

In fact, we shall see that some of the restrictions on LOCC
copying are, if we wish to use entanglement efficiently, more
severe than those on LOCC state discrimination. This will turn out
to be a consequence of the fact that, when copying states by LOCC,
there are certain factors we must take into account that do not
apply to LOCC state discrimination. In LOCC state discrimination,
the original state is typically destroyed. This is of no concern,
since we only wish to know the state, not preserve it. However, in
copying the state, not only do we wish to preserve the original
state, we also wish to imprint it onto another system initialised
in a `blank' state. If we restrict ourselves to performing LOCC
copying and the states we wish to copy are entangled, then the
blank state must be entangled also.  If this were not the case,
then the copying procedure would create entanglement, which is
well known to be impossible under LOCC \cite{ent,Plenio V 98}.

It was recently discovered by Ghosh et al \cite{Ghosh}, in an
independent work, that some sets of orthogonal, maximally
entangled states can be copied by LOCC and with a maximally
entangled blank state. These authors considered LOCC copying of
Bell states. The Bell states, which are maximally entangled states
of two qubits, each contain one ebit of entanglement. These
authors showed that LOCC copying of any two Bell states is
possible with a blank state containing one ebit of entanglement.
They found, however, that to copy all four Bell states requires
one further ebit of entanglement.  This is still less than the two
further ebits that would be required to perform an arbitrary
operation on the  four qubits comprising the states
$|{\psi}_{j}{\rangle}$ and $|b{\rangle}$ by LOCC
\cite{Eisert,CGB,Collins}.

In this paper, we obtain numerous results which relate to the
problem of copying pure, bipartite, orthogonal, entangled states
by LOCC. Throughout, we are interested in making perfect copies
deterministically.  In section \ref{sec2}, we set up the copying
problem in general terms. In doing so, we acknowledge the fact
that an LOCC operation may, in principle, involve an unlimited
number of rounds of classical communication.  As such, the
operation may become unwieldy in formal terms. Rather than deal
with this possibility directly, we take an alternative approach
based on the fact that LOCC operations form a subset of the set of
separable operations. The form of a general separable operation is
well known and more convenient to work with.

For the reasons we gave above, the LOCC copying procedure must use
an entangled blank state.  Entanglement is a precious resource in
quantum information processing.  Consequently, it is highly
desirable that entanglement is used efficiently and if, at all
possible, conserved by the operation. To investigate this matter
fully, we require a measure of entanglement. The problem of
quantifying entanglement is central to quantum information theory.
For pure, bipartite states in the asymptotic limit, where many
copies of the states are available, a unique measure of
entanglement can be provided \cite{Concentration,Popescu}. This is
the entropy of entanglement. However, in the scenario considered
in this paper, we only have one copy of each of the states to be
processed: the state to be copied and the blank state.

In this `one-shot' scenario, there exist pairs of incomparable
states, for which one cannot unambiguously decide whether the
entanglement of one state is greater than, less than or equal to
that of the other. Ideally, we would like the entanglement of the
blank state to equal that of the most entangled of the states to
be copied, as this would represent the most efficient use of
entanglement. However, our desire to use entanglement efficiently
leads us to, in general, account for the possibility of
incomparability of the blank state and some of the states to be
copied.

We show that when all of the states to be copied are non-maximally
entangled, accounting for this possible incomparability leads to
scenarios where, although the LOCC copying procedure is possible,
the blank state cannot be directly transformed into the state to
be copied by LOCC. Instead, the original copy of the state serves
as an entanglement catalyst \cite{Catalysis} which facilitates the
copying procedure. This point is illustrated in the simple case
where we wish to copy just one state.

In section \ref{sec3}, we analyse in detail the problem of locally
copying $N$ orthogonal, entangled states with $D$ dimensional
subsystems.  The blank state is also taken to be an entangled
state whose subsystems are D dimensional.  We focus in particular
on the situation where one of the states to be copied is maximally
entangled. This simplifies the problem in many respects. Firstly,
the possibility of catalytic copying, with its attendant
complications, does not arise, since a maximally entangled state
cannot serve as an entanglement catalyst \cite{Catalysis}.
Consequently, the blank state must be maximally entangled also.
Secondly, we show that the local Kraus operators for a separable
copying operation must be proportional to unitary operators if
they are to copy a maximally entangled state. This is very
helpful, since any separable operation whose Kraus operators have
this property can be performed by LOCC. Indeed, we find that we
may, without loss of generality, take the entire copying operation
to consist of just two local unitary operations, with one being
carried out by each party, and no classical communication.  This
implies that if one of the states to be copied is maximally
entangled, then they must all be maximally entangled. We then use
the convenient form of these operators to obtain a general
necessary and sufficient condition for LOCC copying of $N$ $D$
dimensional maximally entangled states with a maximally entangled
blank state.

This condition is difficult to solve for arbitrary $N$ and $D$.
However, it can be solved exactly for $N=2$ and all $D$. In
section \ref{sec4}, we present this solution in detail and
describe a number of its consequences. In particular, we find that
for $D=2,3$, any pair of maximally entangled, bipartite pure
states can be copied using the same LOCC operation and a maximally
entangled blank state. However, we also show that for any $D$
which is not prime, there exist such pairs for which this copying
operation is impossible.  We conclude in section \ref{sec5} with a
discussion of our results.

\section{The problem of LOCC copying}
\label{sec2}
\renewcommand{\theequation}{2.\arabic{equation}}
\setcounter{equation}{0}
\subsection{General considerations}

\begin{figure}
\begin{center}
\epsfxsize6cm \centerline{\epsfbox{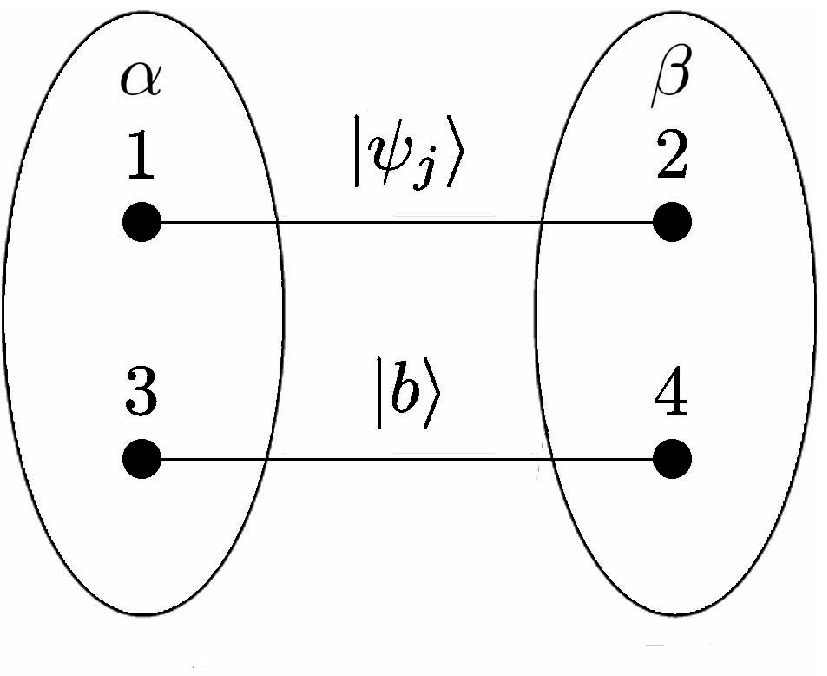}}
\end{center}
\caption{Depiction of the scenario considered in this paper.
Laboratories ${\alpha}$ and ${\beta}$ are spatially separated.
These laboratories contain the pairs of particles (1,3) and (2,4)
respectively. Particles 1 and 2 are initially prepared in one of
the entangled states $|{\psi}_{j}{\rangle}$.  Particles 3 and 4
are initially prepared in the entangled blank state $|b{\rangle}$.
The aim is to perform the copying transformation in Eq.
\eqref{trans1} by LOCC.}
 \label{figure1}
\end{figure}

Let us consider the following scenario, depicted in Figure
\ref{figure1}. We have two parties, Alice and Bob, occupying
spatially separated laboratories ${\alpha}$ and ${\beta}$
respectively. Alice and Bob each have two $D$ dimensional quantum
systems. Alice's systems will be labelled 1 and 3 while Bob's will
be labelled 2 and 4.  Associated with each of these systems is a
copy of the $D$ dimensional Hilbert space ${\cal H}$. The tensor
product Hilbert spaces of Alice's and Bob's pairs will be denoted
by ${\cal H}_{\alpha}$ and ${\cal H}_{\beta}$ respectively. Alice
and Bob also possess ancillary quantum systems enabling them to
carry out arbitrary local quantum operations.  They also share a
two-way classical channel allowing unlimited classical
communication between them.

Consider now a set of entangled, bipartite, pure states
$\{|{\psi}_j{\rangle}\}$, where $j{\in}\{1,...,N\}$.  Throughout
this article, when $N>1$, we shall take the $|{\psi}_{j}{\rangle}$
to be orthogonal.  This implies that, without the LOCC
restriction, the states could be perfectly copied.  Particles 1
and 2 are initially prepared in one of these states although Alice
and Bob do not know which one. Particles 3 and 4 are initially
prepared in the known, bipartite, blank state $|b {\rangle}$.
Alice and Bob aim to perform the transformation
\begin{equation}
\label{trans1} |{\psi}_j^{12}{\rangle}{\otimes} |b^{34}{\rangle}
\rightarrow |{\psi}_j^{12}{\rangle}{\otimes}
|{\psi}_j^{34}{\rangle}
\end{equation}
by LOCC.  Here, the superscripts indicate the particles that have
been prepared in each state.

General quantum state transformations are described using the
quantum operations formalism \cite {Kraus,Chuang}. A quantum
operation on a quantum system with Hilbert space ${\cal S}$ is
represented mathematically by a completely positive, linear, trace
non-increasing map from the set of linear operators on ${\cal S}$
to itself (when the input and output Hilbert spaces are identical,
which is the case in the present context.) Let us denote such a
map by ${\cal E}$ and consider a quantum system whose initial
state is described by a density operator ${\rho}$. This map
transforms the density operator according to
\begin{equation}
\rho\rightarrow \frac{{\cal E}(\rho)}{\mathrm{Tr}({\cal
E}(\rho))}.
\end{equation}

A particularly useful representation of quantum operations is the
operator-sum representation:
\begin{equation}\label{kr}
{\cal E}(\rho)=\sum_{k=1}^{K} F_k\rho F_k^{\dagger},
\end{equation}
where $K$ is some positive integer.  For ${\cal E}$ to be a
physically realisable quantum operation, the $F_{k}$, which are
known as the Kraus operators, must be linear operators that
satisfy
\begin{equation}
\label{res}
\sum_{k=1}^{K} F^{\dagger}_{k} F_{k}\leq{\id},
\end{equation}
where ${\id}$ is the identity operator on ${\cal S}$.   The
equality holds when the map is trace preserving for all states, in
which case the quantum operation is deterministic for all states.
If the operation is not trace preserving for a particular initial
state, then it can only be implemented with probability equal to
the trace of the final state. Whether or not the operation has
been implemented can be always be determined in principle, and
this may be viewed as a generalised measurement. More generally,
any experiment implements a trace preserving sum of trace
non-increasing quantum operations.  The operation that has
actually been carried out can always, in principle, be determined,
and it formally corresponds to a particular outcome of a
generalised measurement.

There are many particular kinds of quantum operation of special
interest. In the present context, two kinds are particularly
important. These are the separable operations \cite{Rains} and the
LOCC procedures. In a separable operation acting on two systems in
spatially separated laboratories ${\alpha}$ and ${\beta}$, the
$F_{k}$ may be written as
\begin{equation}
\label{sepkraus}
 F_{k}=A_{k}{\otimes}B_{k}.
\end{equation}
Here, $A_{k}$ and $B_{k}$ are local Kraus operators acting on
${\cal H}_{\alpha}$ and ${\cal H}_{\beta}$ respectively.  In this
context, we may refer to the $F_{k}$ as the global Kraus
operators.

LOCC procedures are sequences of trace preserving local quantum
operations carried out in the individual laboratories,
interspersed with rounds of classical communication.  The
information received at each laboratory is used to control the
subsequent local operation at the same location. The number of
rounds of classical communication can be arbitrarily large and,
consequently, LOCC procedures can be difficult to work with.
However, the set of such procedures is a subset of the set of
separable operations.  It follows that separability is only a
necessary condition for a quantum operation to be implementable by
LOCC \cite{ent}. It is not sufficient. Still, the fact that the
global Kraus operators for separable operations have the simple
form shown in Eq. \eqref{sepkraus} often makes such operations a
useful starting point for investigating problems relating to LOCC.
See, for example, \cite{USDLOCC}.

\subsection{Catalytic copying}

Due to the limitations on the LOCC manipulation of entanglement,
it is a non-trivial matter to determine the set of blank states
which enable one to copy, by LOCC, even a single, known state
$|{\psi}{\rangle}$. In principle, the conditions under which this
is possible can be obtained using Nielsen's theorem
\cite{Nielsen}. This result specifies the conditions under which
one pure, bipartite, entangled state can be transformed into
another by deterministic LOCC.

Nielsen's theorem involves the concept of majorization, which we
will briefly review.   Consider two real, $R$-component vectors
$v=(v_{1},{\ldots},v_{R})$ and $w=(w_{1},{\ldots},w_{R})$.
 Furthermore, let $v^{\downarrow}$ and $w^{\downarrow}$ be the
 vectors obtained from $v$ and $w$ by arranging their components
 in non-increasing order.  The vector $w$ is said to majorize the
 vector $v$ if
\begin{equation}
\sum_{i=1}^{r}v_{i}^{\downarrow}\leq\sum_{i=1}^{r}w_{i}^{\downarrow},
\end{equation}
for all $r{\in}\{1,{\ldots},R\}$ and with the equality holding for
$R=r$. This majorization relation is usually written as
$w{\succ}v$ or $v{\prec}w$.

Consider now two pure bipartite states $|{\phi}_{1}{\rangle}$ and
$|{\phi}_{2}{\rangle}$.  These may be written in Schmidt
decomposition form as
$|{\phi}_{s}{\rangle}=\sum_{i=1}^{R}\sqrt{{\lambda}_{si}}|x_{si}{\rangle}{\otimes}|y_{si}{\rangle}$,
where $s{\in}\{1,2\}$ and where the maximum subsystem Hilbert
space dimension is $R$. The Schmidt vectors
${\lambda}_{s}=({\lambda}_{s1},{\ldots},{\lambda}_{sR})$ may,
without loss of generality, be taken to have real, non-negative
components.

Nielsen's theorem states that $|{\phi}_{1}{\rangle}$ can be
transformed by deterministic LOCC into $|{\phi}_{2}{\rangle}$ if
and only if
\begin{equation}
\label{maj}
{\lambda}_{1}{\prec}{\lambda}_{2}.
\end{equation}

Returning to problem of LOCC copying, let the states
$|{\psi}{\rangle}$ and $|b{\rangle}$ have the Schmidt vectors
${\lambda}_{\psi}$ and ${\lambda}_{b}$.  We wish to implement the
transformation
\begin{equation}
\label{transingle} |{\psi}^{12}{\rangle}{\otimes} |b^{34}{\rangle}
\rightarrow |{\psi}^{12}{\rangle} {\otimes}|{\psi}^{34}{\rangle},
\end{equation}
by LOCC.  Nielsen's theorem implies that this will be possible if
and only if
\begin{equation}
\label{trump1}
{\lambda}_{\psi}{\otimes}{\lambda}_{b}{\prec}{\lambda}_{\psi}{\otimes}{\lambda}_{\psi}.
\end{equation}
Clearly, this copying transformation will be possible if the
transformation $|b{\rangle}{\rightarrow}|{\psi}{\rangle}$ is
possible by LOCC, i.e., if ${\lambda}_{b}{\prec}{\lambda}_{\psi}$.
However, what if $|b{\rangle}{\rightarrow}|{\psi}{\rangle}$ is
impossible by LOCC? When $|b{\rangle}$ cannot be transformed into
$|{\psi}{\rangle}$ by deterministic LOCC, there appear at first
sight to be two cases to consider, corresponding to whether or not
$|{\psi}{\rangle}{\rightarrow}|b{\rangle}$ is possible by LOCC.
However, we shall now show that the possibility of the LOCC
transformation $|{\psi}{\rangle}{\rightarrow}|b{\rangle}$, when
combined with our assumptions that the copying transformation in
Eq. \eqref{transingle} is possible by LOCC and that the
transformation $|{\psi}{\rangle}{\rightarrow}|b{\rangle}$ isn't,
leads to a contradiction.

To do this, it is useful to introduce another relation between two
vectors, the trumping relation. Consider two real vectors $v$ and
$w$.  If there exists a real vector $u$ such that
\begin{equation}
\label{trump2} u{\otimes}v{\prec}u{\otimes}w,
\end{equation}
then we say that `$w$ trumps $v$' and write this relation as
$v{\prec}_{T}w$ or $w{\succ}_{T}v$.  From Eq. \eqref{trump1}, we
clearly see that
\begin{equation}
\label{trump3} {\lambda}_{b}{\prec}_{T}{\lambda}_{\psi}.
\end{equation}
The trumping relation is weaker than the majorization relation:
that is, if $v{\prec}w$ then $v{\prec}_{T}w$, but not necessarily
vice versa.  We are assuming that
$|{\psi}{\rangle}{\rightarrow}|b{\rangle}$ is possible by LOCC,
which implies that ${\lambda}_{\psi}{\prec}{\lambda}_{b}$.
Therefore,
\begin{equation}
\label{trump4}
{\lambda}_{\psi}{\prec}_{T}{\lambda}_{b}.
\end{equation}

We shall now use the following theorem due to Jonathan and Plenio
\cite{Catalysis}: if $v{\prec}_{T}w$ and $w{\prec}_{T}v$, then
$v^{\downarrow}=w^{\downarrow}$.  Combining this result with Eqs.
\eqref{trump3} and \eqref{trump4}, we see that
${\lambda}_{\psi}^{\downarrow}={\lambda}_{b}^{\downarrow}$.  When
this is so, it follows that ${\lambda}_{b}{\prec}{\lambda}_{\psi}$
and, by Nielsen's theorem, that the transformation
$|b{\rangle}{\rightarrow}|{\psi}{\rangle}$ is actually possible by
LOCC, which contradicts our premise.

The remaining possibility is that both
$|b{\rangle}{\rightarrow}|{\psi}{\rangle}$ and
$|{\psi}{\rangle}{\rightarrow}|b{\rangle}$ are impossible to
perform by LOCC. When this is the case, the states
$|{\psi}{\rangle}$ and $|b{\rangle}$ are said to be incomparable.
Even though incomparable states cannot be transformed into each
other by LOCC, there is the possibility that the transformation in
\eqref{transingle} is possible.  When this is so,
$|{\psi}^{12}{\rangle}$, which is unchanged by the copying
procedure, is said to act as a catalyst for the transformation
$|b^{34}{\rangle}{\rightarrow}|{\psi}^{34}{\rangle}$.

 The problem  of finding, for a general state
 $|{\psi}{\rangle}$, the set of blank states $|b{\rangle}$ for
 which $|{\psi}{\rangle}$ can be copied by entanglement catalysis is a challenging task.  This is due to the fact that no
analytical way of  ordering the Schmidt coefficients of a general
tensor product of two states has yet been discovered.
Nevertheless, by numerical methods, one can easily check for
particular states whether or not the majorization relation in Eq.
\eqref{maj} is satisfied. One can then search for pairs of pure,
bipartite entangled states such that one cannot be transformed
into another directly but for which the transformation is possible
with a catalyst.  A specific example of catalytic copying, which
we obtained in this way, is as follows. Consider the case of $D=5$
and a state $|{\psi}{\rangle}$ with Schmidt coefficients
$\sqrt{0.39},\sqrt{0.26},\sqrt{0.18},\sqrt{0.17}$ and $0$.
Consider also a blank state $|b{\rangle}$ with Schmidt
coefficients $\sqrt{0.32},\sqrt{0.28},\sqrt{0.24},\sqrt{0.085}$
and $\sqrt{0.075}$. For these two states, one can readily verify
using Nielsen's theorem that the transformation
$|b{\rangle}\rightarrow|{\psi}{\rangle}$ is impossible by LOCC
while the transformation $|{\psi}^{12}{\rangle}{\otimes}|b^{34}
{\rangle}\rightarrow
|{\psi}^{12}{\rangle}{\otimes}|{\psi}^{34}{\rangle}$ can be
carried out this way.

The main focus of this paper is on LOCC copying of multiple
quantum states with efficient use of entanglement.  Even for a
single, known state, the problem is complicated by the possibility
of catalytic copying as we have just demonstrated.  To generalise
this to multiple states, we would require an understanding of
multi-state catalytic entanglement transformations, about which
little, if anything, is currently known. Fortunately, there is a
large class of states sets that we can consider for which the
issue of catalysis does not arise. These are sets where at least
one of the states to be copied is maximally entangled. Their
preferential status is a consequence of the fact that maximally
entangled states cannot serve as catalysts for pure, bipartite
entanglement transformations \cite{Catalysis}.  Such sets will be
the focus of our attention for the remainder of this paper.

\section{LOCC copying of a pure orthogonal set including a maximally entangled state}
\label{sec3}
\renewcommand{\theequation}{3.\arabic{equation}}
\setcounter{equation}{0}
\subsection{Form of the local Kraus
operators}

Returning to the problem of locally copying the $N$ states
$|{\psi}_{j}{\rangle}$, recall that we require the copying
operation to be separable. This implies that the global Kraus
operators will have the form shown in Eq. \eqref{sepkraus}, where
the $A^{13}_{k}$ and $B^{24}_{k}$ act on ${\cal H}_{\alpha}$ and
${\cal H}_{\beta}$ respectively. In terms of these operators, the
copying transformation will have the form
\begin{equation}
\label{trans2} A_{k}^{13}{\otimes} B_{k}^{24}
|{\psi}_{j}^{12}{\rangle}{\otimes} |b^{34}{\rangle}=\sigma_{jk}
|{\psi}_{j}^{12}{\rangle}{\otimes} |{\psi}_{j}^{34}{\rangle}.
\end{equation}
Here, the superscripts on the operators indicate the particles on
which they act.  Also, the $\sigma_{jk}$ are some complex
coefficients that satisfy $\sum_{k=1}^{K}|\sigma_{jk}|^{2}=1$.

Separability of the copying operation is, as we have noted above,
only a necessary and not a sufficient condition for LOCC copying.
However the combination of the separability condition with
specific features relating to particular sets of states can lead
us to exact necessary and sufficient conditions for LOCC copying.
The remainder of this paper is devoted to investigating the LOCC
copying problem for a class of such sets.  These are sets where at
least one of the states to be copied is maximally entangled.

For the sake of definiteness, let the state $|{\psi}_{1}{\rangle}$
be maximally entangled. It is known \cite{Catalysis} that a
maximally entangled state cannot serve as a catalyst.  Therefore,
the transformation $|b{\rangle}{\rightarrow}|{\psi}_{1}{\rangle}$
must be possible by LOCC.  Since we are restricting ourselves to
blank states of a pair of $D$ dimensional particles, it follows
from Nielsen's theorem that the blank state is necessarily
maximally entangled also.

This section is devoted to determining the conditions under which
the $|{\psi}_{j}{\rangle}$ can be copied by LOCC when both
$|{\psi}_{1}{\rangle}$ and the blank state $|b{\rangle}$ are
maximally entangled.  In the first part of this section, we will
see how the requirements of our operation have interesting
implications for the form of the local Kraus operators in Eq.
\eqref{trans2}.  We will then obtain the general necessary and
sufficient conditions under which our desired operation is
physically possible.

To begin, let $\{|x_{i}{\rangle}\}$ be an orthonormal basis for
the single particle Hilbert space ${\cal H}$. We will frequently
work with the following reference maximally entangled state in
${\cal H}^{{\otimes}2}$:
\begin{equation}
\label{max} |{\psi}_{max}^{rs}{\rangle}
=\frac{1}{\sqrt{D}}\sum_{i=1}^{D}
|x_{i}^{r}{\rangle}{\otimes}|x_{i}^{s}{\rangle}.
\end{equation}
We will also frequently encounter the product states
$|x^{r}_{i}{\rangle}{\otimes}|x^{s}_{j}{\rangle}$, for particles
$r,s$ where $r,s{\in}\{1,{\ldots},4\}$.  As such, it is convenient
to adopt a simpler notation for these states.  Define
\begin{equation}
|X_{\mu}^{rs}{\rangle}=|x_{i}^{r}{\rangle}{\otimes}|x_{j}^{s}{\rangle},
\end{equation}
where ${\mu}={\mu}(i,j){\in}\{1,{\ldots},D^{2}\}$.  Each value of
${\mu}$ must correspond to unique values of $i$ and $j$.  This can
be achieved, for example, by letting ${\mu}=i+D(j-1)$ with
$i,j\in\{1,\ldots,D\}$.  More generally, we will use Greek
subscripts to index elements of this basis according to the same
formula as for ${\mu}$.

The fact that the state $|{\psi}_{1}{\rangle}$ is maximally
entangled implies that there exists a unitary operator $U_{1}$ on
${\cal H}$ such that
\begin{equation}
\label{MM} |{\psi}_1^{12}{\rangle}= (U^{1}_1{\otimes}{\id}^{2})
|{\psi}^{12}_{max}{\rangle}.
\end{equation}
When the particle pair (3,4) is in this state, we replace the
superscripts 1 and 2 with 3 and 4 respectively.

The blank state $|b{\rangle}$ is also maximally entangled, so
there exists a unitary operator $U_{b}$ on ${\cal H}$ such that
\begin{equation}
\label{bdef} |b^{34}{\rangle}=
(U^{3}_b{\otimes}{\id}^{4})|{\psi}^{34}_{max}{\rangle}.
\end{equation}

We now proceed to show that, without loss of generality, the
$A^{13}_{k}$ and $B^{24}_{k}$ in Eq. \eqref{trans2} may be taken
to be, up to multiplicative coefficients, unitary.  To do this, we
note that the most general LOCC procedure consists of an
arbitrarily long sequence of local operations in Alice's and Bob's
laboratories interspersed with rounds of classical communication.
The entire LOCC operation is initiated by one party. For the sake
of definiteness, and without loss of generality, let this party be
Alice. Alice implements a deterministic local operation on her
system. This operation, which is trace preserving, may be a sum of
trace non-increasing operations in which Alice obtains (classical)
information about which of these operations was carried out. The
entire operation is then a generalised measurement. If it is, then
the measurement result is communicated to Bob. Upon receiving
this, Bob implements a local operation corresponding to this
result. He then communicates a description of his operation to
Alice (if she does not already know the operation he will perform
given the classical information she sent him) together with any
measurement results and the process can repeat an arbitrarily
large number of times.

The crucial point is the fact that if Alice and Bob begin with the
state $|{\psi}_{1}^{12}{\rangle}{\otimes}|b^{34}{\rangle}$, which
is a maximally entangled state of the pairs (1,3) and (2,4), then
the LOCC copying procedure will produce the state
$|{\psi}_{1}^{12}{\rangle}{\otimes}|{\psi}_{1}^{34}{\rangle}$,
which is also a maximally entangled state of these pairs of
particles. No LOCC procedure can transform a maximally entangled
state into a non-maximally entangled state, and then into another
maximally entangled state.  It follows that each step in their
LOCC copying procedure can do no more than transform one maximally
entangled state of these pairs of particles into another.  So, let
$|{\chi}_{1}{\rangle}$ and $|{\chi}_{2}{\rangle}$ be maximally
entangled states of the pairs (1,3) and (2,4). We may write these
states as
\begin{equation}
\label{phir}
|{\chi}_{r}{\rangle}=(V^{13}_{r}{\otimes}{\id}^{24})|{\psi}^{12}_{max}{\rangle}{\otimes}|{\psi}^{34}_{max}{\rangle},
\end{equation}
where $r{\in}\{1,2\}$ and the $V^{13}_{r}$ are unitary operators
acting on ${\cal H}_{\alpha}$.  We will now investigate the
properties of a local operation in one laboratory that transforms
$|{\chi}_{1}{\rangle}$ into $|{\chi}_{2}{\rangle}$.   For the sake
of definiteness, we let this operation be carried out by Alice in
her laboratory ${\alpha}$.  The following argument applies equally
well if the operation were to be carried out by Bob. Alice carries
out a local operation, which we shall denote by ${\cal E}^{13}$.
This takes the form of a completely positive, linear, trace
non-increasing map on the space of linear operators on ${\cal
H}_{\alpha}$. Interpreting this operation as corresponding to a
generalised measurement outcome, whose probability is $p$ for the
initial state $|{\chi}_{1}{\rangle}$, this operation must produce
the state $|{\chi}_{2}{\rangle}$ according to
\begin{equation}
\label{E13} {\cal
E}^{13}{\otimes}{\id}^{24}(|{\chi}_{1}{\rangle}{\langle}{\chi}_{1}|)=p|{\chi}_{2}{\rangle}{\langle}{\chi}_{2}|.
\end{equation}
Let us now define the following operation on particles 1 and 3
whose action on an arbitrary density operator ${\rho}^{13}$ of
these particles is
\begin{equation}
\label{tildeE131} \tilde{\cal
E}^{13}({\rho}^{13})=V_{2}^{{\dagger}13}{\cal
E}^{13}(V_{1}^{13}{\rho}^{13}V_{1}^{{\dagger}13})V_{2}^{13}.
\end{equation}
From Eqs. \eqref{phir}, \eqref{E13} and \eqref{tildeE131} we see
that
\begin{eqnarray}
\label{tildeE132} &&\tilde{\cal
E}^{13}{\otimes}{\id}^{24}(|{\psi}^{12}_{max}{\rangle}{\langle}{\psi}^{12}_{max}|{\otimes}|{\psi}^{34}_{max}{\rangle}{\langle}{\psi}^{34}_{max}|)
\nonumber
\\&=&p|{\psi}^{12}_{max}{\rangle}{\langle}{\psi}^{12}_{max}|{\otimes}|{\psi}^{34}_{max}{\rangle}{\langle}{\psi}^{34}_{max}|.
\end{eqnarray}
We will now proceed to show that the above transformation implies
that
\begin{equation}
\label{EID} \tilde{\cal
E}^{13}({\cdot})=p{\id}^{13}({\cdot}){\id}^{13}.
\end{equation}
To do so, let us expand Eq. \eqref{tildeE132} in terms of the
$|X_{\mu}{\rangle}$ basis states, which gives
\begin{eqnarray}
&&\sum_{{\mu},{\nu}=1}^{D^{2}}\tilde{\cal
E}^{13}(|X^{13}_{\mu}{\rangle}{\langle}X^{13}_{\nu}|){\otimes}|X^{24}_{\mu}{\rangle}{\langle}X^{24}_{\nu}|
\nonumber \\
&=&p\sum_{{\mu},{\nu}=1}^{D^{2}}|X^{13}_{\mu}{\rangle}{\langle}X^{13}_{\nu}|{\otimes}|X^{24}_{\mu}{\rangle}{\langle}X^{24}_{\nu}|,
\end{eqnarray}
where we have omitted the overall factor of $1/D^{2}$.  Acting on
the (2,4) states to the left with ${\langle}X^{24}_{\gamma}|$ and
to the right with $|X^{24}_{\delta}{\rangle}$ and making use of
their orthonormality, we obtain
\begin{equation}
\label{transE1} \tilde{\cal
E}^{13}(|X^{13}_{\gamma}{\rangle}{\langle}X^{13}_{\delta}|)=p|X^{13}_{\gamma}{\rangle}{\langle}X^{13}_{\delta}|.
\end{equation}
An arbitrary linear operator ${\Omega}$ acting on ${\cal
H}^{{\otimes}2}$ may be written as
\begin{equation}
{\Omega}=\sum_{{\gamma},{\delta}=1}^{D^{2}}{\omega}_{{\gamma}{\delta}}|X_{\gamma}{\rangle}{\langle}X_{\delta}|,
\end{equation}
having the matrix elements ${\omega}_{{\gamma}{\delta}}$ in the
$|X_{\gamma}{\rangle}$ basis.  From Eq. \eqref{transE1}  and the
linearity of $\tilde{\cal E}^{13}$, it readily follows that
\begin{equation}
\label{transE2} \tilde{\cal E}^{13}({\Omega}^{13})=p{\Omega}^{13}.
\end{equation}
Since this is true for any linear operator ${\Omega}$ on ${\cal
H}^{{\otimes}2}$, we require that Eq. \eqref{EID} is true.

Combining this with Eq. \eqref{tildeE131}, we see that Alice's
operation has the form
\begin{equation}
\label{tildeE} \tilde{\cal
E}^{13}({\rho}^{13})=p(V_{2}V_{1}^{\dagger})^{13}{\rho}^{13}(V_{1}V_{2}^{\dagger})^{13}.
\end{equation}
In this local operation, the Kraus operators may be taken to be
$\sqrt{p}(V_{2}V_{1}^{\dagger})^{13}$, which are clearly
proportional to unitary operators.  Furthermore, Alice's overall
local Kraus operators $A^{13}_{k}$ are simply the products of the
local Kraus operators corresponding to the elementary steps she
carries out in the entire LOCC procedure. These must also be
proportional to unitary operators, since the product of any number
of unitary operators is also a unitary operator. Clearly, the
above argument also applies if the elementary step is carried out
by Bob.  We are therefore led to the following conclusion: the
local Kraus operators $A^{13}_{k}$ and $B^{24}_{k}$ for the entire
LOCC procedure are, up to overall multiplicative coefficients,
unitary. These coefficients are real and non-negative since they
are, from our above definition of the elementary step local Kraus
operators, products of the square roots of probabilities.  We may
then write
\begin{eqnarray}
A^{13}_{k}&=&f_{k}\tilde{A}^{13}_{k}
\\B^{24}_{k}&=&g_{k}\tilde{B}^{24}_{k}.
\end{eqnarray}
Here, $\tilde{A}^{13}_{k}$ and $\tilde{B}^{24}_{k}$ are unitary
operators on ${\cal H}_{\alpha}$ and ${\cal H}_{\beta}$
respectively and the $f_{k},g_{k}$ are the real, non-negative
coefficients which satisfy
\begin{equation}
\label{coeffprob} \sum_{k=1}^{K}(f_{k}g_{k})^{2}=1,
\end{equation}
as a consequence of Eq. \eqref{res} and the fact that our LOCC
procedure is trace preserving.

This has several important consequences that we can take advantage
of.  The first is the fact that any separable quantum operation
whose local Kraus operators have this property can be carried out
by LOCC.  This can be done in the following way. At one of the
laboratories, say ${\alpha}$, a random variable $Y$ with $K$
possible values $y_{k}$ and probability distribution
$p_{k}=(f_{k}g_{k})^{2}$ is generated. On obtaining the result
$y_{k}$, Alice carries out the local unitary operation
$\tilde{A}^{13}_{k}$.  She also communicates the value of $Y$ to
Bob, who then proceeds to implement the transformation
$\tilde{B}^{24}_{k}$.

The fact that the global Kraus operators $F_{k}$ are, up to
multiplicative coefficients, unitary implies that each one can be
implemented deterministically.  Furthermore, they must each carry
out the desired LOCC copying transformation, for each of the
states to be copied. Otherwise, the final state would be mixed.
This implies that a necessary and sufficient condition for
implementing the copying transformation is that the copying
procedure can be implemented by a single global Kraus operator
$F=A^{13}{\otimes}B^{24}$, where $A^{13}$ and $B^{24}$ are
unitary. When this is the case, the complex coefficients
${\sigma}_{jk}$ in Eq. \eqref{trans2}, where we may drop the index
$k$, have unit modulus.  Implementing these observations, Eq.
\eqref{trans2} becomes
\begin{equation}
\label{trans3}
A^{13}{\otimes}B^{24}|{\psi}^{12}_{j}{\rangle}{\otimes}|b^{34}{\rangle}=e^{i{\theta}_{j}}|{\psi}^{12}_{j}{\rangle}{\otimes}|{\psi}_{j}^{34}{\rangle},
\end{equation}
for some angles ${\theta}_{j}$.   The fact that $A$ and $B$ are
unitary implies that the states $|{\psi}_{j}{\rangle}$ must all be
maximally entangled.  The reason for this is that, if any
non-maximally entangled state $|{\psi}_{j}{\rangle}$ could be
perfectly copied, then particles 3 and 4, initially prepared in
the maximally entangled state $|b{\rangle}$, would be left in the
non-maximally entangled state $|{\psi}_{j}{\rangle}$.  This is
impossible to achieve with a pair of local unitary operators.

In the remainder of this section, we shall use the above findings
to obtain a general necessary and sufficient condition for LOCC
copying, with a maximally entangled blank state, of the states
$|{\psi}_{j}{\rangle}$ when they are orthonormal and maximally
entangled.

\subsection{Condition for LOCC copying}

We saw above that, if $|{\psi}_{1}{\rangle}$ is maximally
entangled, then the $|{\psi}_{j}{\rangle}$ are all maximally
entangled. Consequently, we may write all of these states in the
same form as we did for $|{\psi}_{1}{\rangle}$ in Eq. \eqref{MM},
that is, as
\begin{equation}
\label{psiju} |{\psi}_{j}^{12}{\rangle}=
(U_{j}^{1}{\otimes}{\id}^{2})|{\psi}_{max}^{12}{\rangle},
\end{equation}
for some unitary operators $U_{j}$ on ${\cal H}$.  Again, when
considering the particle pair (3,4) in one of these states, we
will change the superscripts 1 and 2 to 3 and 4 respectively.

Let us now define the following unitary operators on ${\cal
H}_{\alpha}$:
\begin{equation}
\label{cj} C_{j}^{13}=(U^{{\dagger}1}_{j}{\otimes}
U^{{\dagger}3}_{j})A^{13}(U_{j}^{1}{\otimes}U^{3}_{b}).
\end{equation}
With a small amount of algebra, it is easily seen that Eq.
\eqref{trans3} is equivalent to
\begin{equation}
C^{13}_{j}{\otimes}B^{24}|{\psi}^{12}_{max}{\rangle}{\otimes}|{\psi}_{max}^{34}{\rangle}=e^{i{\theta}_{j}}|{\psi}^{12}_{max}{\rangle}{\otimes}|{\psi}_{max}^{34}{\rangle}.
\end{equation}
In terms of the two-particle basis set $\{|X_{\mu}{\rangle}\}$,
this can be written as
\begin{equation}
C^{13}_{j}{\otimes}B^{24}\sum_{\mu=1}^{D^2}|
X^{13}_{\mu}{\rangle}{\otimes} | X^{24}_{\mu}{\rangle}=
e^{i{\theta}_{j}}\sum_{\mu=1}^{D^2}|
X^{13}_{\mu}{\rangle}{\otimes} | X^{24}_{\mu}{\rangle}.
\end{equation}
Notice that the $|X^{13}_{\nu}{\rangle}{\otimes}|
X^{24}_{\tau}{\rangle}$ form a basis for the total Hilbert space
${\cal H}_{\alpha}{\otimes} {\cal H}_{\beta}$. Acting to the left
throughout with ${\langle}
X^{13}_{\nu}|{\otimes}{\langle}X^{24}_{\tau}|$ we obtain,
\begin{eqnarray}
\label{CBmatrix1} \sum_{\mu=1}^{D^2}{\langle}
X^{13}_{\nu}|C^{13}_{j}|
X^{13}_{\mu}{\rangle}{\langle}X^{24}_{\tau}|B^{24}|
X^{24}_{\mu}{\rangle}&=&e^{i{\theta}_{j}}\sum_{\mu=1}^{D^2}\delta_{\nu\mu}\delta_{\tau\mu} \nonumber \\
&=&e^{i{\theta}_{j}}{\delta}_{{\nu}{\tau}}.
\end{eqnarray}
This can be written as
\begin{equation}
\label{CBmatrix2} C_{j}B^{T}=e^{i{\theta}_{j}}\mathbf{\id}.
\end{equation}
Here, $\mathbf{\id}$ is the identity operator on ${\cal
H}^{{\otimes}2}$ and $T$ denotes the transpose in the
$\{|X_{\mu}{\rangle}\}$ basis. Solving for $B$ and making use of
unitarity, we find that
\begin{equation}
\label{bis} B=e^{i{\theta}_{j}}C_{j}^{*},
\end{equation}
where * denotes complex conjugation in the $\{|X_{\mu}{\rangle}\}$
basis. From this, we see that the operators
$e^{-i{\theta}_{j}}C_{j}$ are independent of $j$. Using the
explicit expression for $C^{13}_{j}$ in Eq. \eqref{cj}, we see
this means that
\begin{eqnarray}
&&e^{-i{\theta}_{j}}(U^{{\dagger}1}_{j}{\otimes}
U^{{\dagger}3}_{j})A^{13}(U_{j}^{1}{\otimes} U_{b}^{3}) \nonumber
\\ &=&e^{-i{\theta}_{j'}}(U^{{\dagger}1}_{j'}{\otimes}
U^{{\dagger}3}_{j'})A^{13}(U_{j'}^{1}{\otimes}U_{b}^{3}),
\end{eqnarray}
for all $j,j'{\in}\{1,{\ldots},N\}$.  Acting throughout to the
left with $U^{1}_{j}{\otimes}U^{3}_{j}$ and to the right with
$U^{{\dagger}1}_{j'}{\otimes} U^{{\dagger}3}_{b}$ we obtain
\begin{eqnarray}
\label{condu}
&&e^{-i{\theta}_{j}}A^{13}[(U_{j}U^{\dagger}_{j'})^{1}{\otimes}
{\id}^{3}] \nonumber
\\ &=&e^{-i{\theta}_{j'}}[(U_{j}U^{\dagger}_{j'})^{1}{\otimes}(
U_{j}U^{\dagger}_{j'})^{3}]A^{13}.
\end{eqnarray}
Prior to proceeding, we shall make a brief digression. From this
point onwards, we will be concerned with operator equations
involving just two particles in a shared entangled state.
Consequently, it will be convenient to drop the particle
superscripts.  We do this because the particles involved will
follow the tensor product ordering convention we established for
such particle pairs in Eq. \eqref{MM} and the subsequent
paragraph. Also, the analysis that follows in the next section
will be quite intricate and will not benefit from unnecessary
notation.

For the sake of notational convenience, define the unitary
operators
\begin{equation}
\label{tjj} T_{jj'}=U_{j}U^{\dagger}_{j'}.
\end{equation}
Using this and the unitarity of $A$, we find that Eq.
\eqref{condu} is equivalent to
\begin{equation}
\label{gencond}
 A(T_{jj'}{\otimes}{\id})A^{{\dagger}}
=e^{i({\theta}_{j}-{\theta}_{j'})}(T_{jj'}{\otimes}T_{jj'}).
\end{equation}
From the above argument, it follows that the existence of a
unitary operator $A$ on ${\cal H}^{{\otimes}2}$ which satisfies
this equation, for some angles ${\theta}_{j}$ and ${\theta}_{j'}$,
is both necessary and sufficient for the existence of an LOCC
copying procedure which, with a maximally entangled blank state
$|b{\rangle}$, copies all of the $|{\psi}_{j}{\rangle}$.

The next section will be devoted to the case of $N=2$.  Prior to
addressing this case, we shall make some further general
observations.  Having defined the operators $U_{j}$ in terms of
the reference maximally entangled state $|{\psi}_{max}{\rangle}$
in Eq. \eqref{psiju}, one might suspect that the $T_{jj'}$ also
make implicit reference to this state.  However, this is not so.
We can, in fact, write these operators solely in terms of the
states to be copied, $|{\psi}_{j}{\rangle}$, and $D$, the
dimensionality of ${\cal H}$. To do so, consider
\begin{equation}
|{\psi}_{j}{\rangle}{\langle}{\psi}_{j'}|=\frac{1}{D}\sum_{i,i'=1}^{D}U_{j}|x_{i}{\rangle}{\langle}x_{i'}|U^{\dagger}_{j'}{\otimes}|x_{i}{\rangle}{\langle}x_{i'}|,
\end{equation}
where we have used Eq. \eqref{max}.  Denoting by `PT' the partial
trace with respect to the second system, we find
\begin{eqnarray}
\label{indep}
D{\times}\mathrm{PT}\left(|{\psi}_{j}{\rangle}{\langle}{\psi}_{j'}|\right)&=&\sum_{i,i'=1}^{D}U_{j}|x_{i}{\rangle}{\langle}x_{i'}|U^{\dagger}_{j'}{\otimes}\mathrm{Tr}(|x_{i}{\rangle}{\langle}x_{i'}|),
\nonumber
\\
&=&\sum_{i=1}^{D}U_{j}|x_{i}{\rangle}{\langle}x_{i}|U^{\dagger}_{j'}=T_{jj'}.
\end{eqnarray}
Here we have used Eq. \eqref{tjj} and the completeness of the
$|x_{i}{\rangle}$.  We see that the copying condition in Eq.
\eqref{gencond} can be expressed solely in terms of the states to
be copied and the dimensionality of the single particle Hilbert
space.

Notice that, from Eq. \eqref{indep}, if we take the full trace of
$|{\psi}_{j}{\rangle}{\langle}{\psi}_{j'}|$ we obtain
\begin{equation}
\label{overlap}
{\langle}{\psi}_{j'}|{\psi}_{j}{\rangle}=\frac{1}{D}\mathrm{Tr}(T_{jj'}).
\end{equation}
It is known from the original no cloning theorem that, for perfect
copying to be possible, we require the states
$|{\psi}_{j}{\rangle}$ and $|{\psi}_{j'}{\rangle}$ to be either
orthogonal or, up to a phase, identical.  It is interesting to see
how this fact also follows from Eq. \eqref{gencond}.  Taking the
full trace throughout Eq. \eqref{gencond} and making use of the
unitarity of $A$, we obtain
\begin{equation}
D\mathrm{Tr}(T_{jj'})=e^{i({\theta}_{j}-{\theta}_{j'})}[\mathrm{Tr}(T_{jj'})]^{2}.
\end{equation}
This is a simple quadratic equation in $\mathrm{Tr}(T_{jj'})$,
whose roots are $0$ and $De^{-i({\theta}_{j}-{\theta}_{j'})}$.
From Eq. \eqref{overlap}, we easily see that these roots
correspond to $|{\psi}_{j}{\rangle}$ and $|{\psi}_{j'}{\rangle}$
being orthogonal and, up to a phase, identical respectively.

The problem of determining when a unitary operator $A$ on ${\cal
H}^{{\otimes}2}$ satisfying Eq. \eqref{gencond} exists appears to
be quite challenging for arbitrary $N$ and $D$. However, for
$N=2$, the problem can be solved exactly for all $D$.  We will
present the detailed solution to this problem and explore some of
its consequences in the next section.

\section{LOCC copying of two orthogonal maximally entangled states}
\label{sec4}
\renewcommand{\theequation}{4.\arabic{equation}}
\setcounter{equation}{0}

\subsection{A spectral copying condition}
From the above discussion, it follows that a necessary and
sufficient condition for LOCC copying of two maximally entangled
states $|{\psi}_{1}{\rangle}$ and $|{\psi}_{2}{\rangle}$ with a
maximally entangled blank state is that there exists a
two-particle unitary operator $A$ which implements the
transformation in Eq. \eqref{gencond} for $j,j'{\in}\{1,2\}$ and
some angles ${\theta}_{1}$ and ${\theta}_{2}$.  Notice from the
definition of the $T_{jj'}$ in Eq. \eqref{tjj} that
$T_{jj}={\id}$, the identity operator on ${\cal H}$. Consequently,
for $j=j'$, Eq. \eqref{gencond} is trivially satisfied by any
unitary operator $A$ and any angles ${\theta}_{j}$. Also, the
equations for $T_{12}$ and $T_{21}$ are simply the Hermitian
adjoints of each other, so if one is true then so is the other. It
follows that for the case of $N=2$, we need only consider one of
these equations.  For the sake of definiteness, we will focus on
the operator $T_{12}$, which we will write simply as $T$. We also
write ${\Delta}{\theta}={\theta}_{1}-{\theta}_{2}$. For suitable
choices of ${\theta}_{1}$ and ${\theta}_{2}$, this can take any
real value. Our condition then becomes
\begin{equation}
\label{2cond1} A(T{\otimes}{\id})A^{\dagger}
=e^{i{\Delta}{\theta}}(T{\otimes}T),
\end{equation}
where ${\id}$ is again the identity operator on ${\cal H}$. We can
simplify this expression further by the removing the phase factor
in the following way: define
\begin{equation}
\label{tildetdef} \tilde{T}=e^{i{\Delta}{\theta}}T.
\end{equation}
Then by simple substitution we find that Eq. \eqref{2cond1} is
equivalent to
\begin{equation}
\label{ttrans}
A(\tilde{T}{\otimes}{\id})A^{\dagger}=\tilde{T}{\otimes}\tilde{T}.
\end{equation}
A unitary operator $A$ satisfying this equation exists if and only
if $\tilde{T}{\otimes}{\id}$ and $\tilde{T}{\otimes}\tilde{T}$
have the same eigenvalues, with the same multiplicities.  So, we
may write our condition for LOCC copying of the two states as
\begin{equation}
\label{specond1}
\mathrm{spec}(\tilde{T}{\otimes}\tilde{T})=\mathrm{spec}(\tilde{T}{\otimes}{\id}),
\end{equation}
where `spec' denotes the spectrum.

Throughout this section, it will be convenient to group the
eigenvalues according to multiplicity.  So, let $M{\leq}D$ be the
number of distinct eigenvalues.  We shall write these as
${\lambda}_{r}$ where $r{\in}\{1,{\ldots},M\}$.  It is easy to see
from Eq. \eqref{specond1} that, for every integer $R{\geq}2$, we
have
\begin{equation}
\label{specond2}
\mathrm{spec}(\tilde{T}^{{\otimes}R})=\mathrm{spec}(\tilde{T}{\otimes}{\id}^{{\otimes}(R-1)}).
\end{equation}
This implies that
\begin{equation}
\label{Rspec}
{\lambda}_{r_{1}}{\lambda}_{r_{2}}{\ldots}{\lambda}_{r_{R}}{\in}\mathrm{spec}(\tilde{T})
\end{equation}
for all $r_{j}{\in}\{1,{\ldots},M\}$ and $j{\in}\{1,{\ldots},R\}$.
To determine which pairs of maximally entangled states can be
simultaneously locally copied with a maximally entangled blank
state, we must find out which unitary operators satisfy Eq.
\eqref{specond1}. The current section will focus on solving this
problem and exploring some of the consequences of its solution.

Prior to giving this solution, we make the following intriguing
observation.  The physical problem of LOCC copying leads to the
mathematical problem expressed in Eq. \eqref{specond1}, where
physical considerations require that $\tilde{T}$ is unitary.
However, if we are interested in this equation from a purely
mathematical perspective, then there is the question of what
properties a general linear operator ${\tilde T}$ must have in
order to solve Eq. \eqref{specond1}. We will now show that the
eigenvalues of any linear operator, if they are all non-zero, must
have unit modulus in order to satisfy  Eq. \eqref{specond1}.

To prove this, we make use of the fact that we may, without loss
of generality, take the ${\lambda}_{r}$ to be arranged in
non-increasing order in terms of their moduli:
\begin{equation}
\label{lambdaineq}
    0<|\lambda_1|{\leq}|\lambda_2|{\leq}{\ldots}{\leq}|\lambda_M|.
\end{equation}
Let us notice that Eq. \eqref{specond1} implies that
${\lambda}_{1}^{2}{\in}\mathrm{spec}(\tilde{T})$. We now assume
that $|\lambda_1|={\min}_{r}\{|{\lambda}_{r}|\}<1$. It immediately
follows that
$|\lambda_1^{2}|=|\lambda_1|^{2}<{\min}_{r}\{|{\lambda}_{r}|\}$
for ${\lambda}_{1}{\neq}0$, contradicting this assumption. Our
assumption must therefore be false.  Similarly, we see that Eq.
\eqref{specond1} implies that
${\lambda}_{M}^{2}{\in}\mathrm{spec}(\tilde{T})$. Let us assume
that $|\lambda_M|={\max}_{r}\{|{\lambda}_{r}|\}>1$. We then obtain
$|\lambda_M^{2}|=|\lambda_M|^2>{\max}_{r}\{|{\lambda}_{r}|\}$,
which also leads to a contradiction.  This argument implies that
$|{\lambda}_{r}|=1$ and leads to the conclusion that the non-zero
eigenvalues must be of the form
\begin{equation}
\label{mod1} {\lambda}_{r}=e^{i{\phi}_{r}}
\end{equation}
for some angles ${\phi}_{r}{\in}[0,2{\pi})$.  Without loss of
generality, we may take these angles to be ordered according to
\begin{equation}
\label{phiinequ1}
0{\leq}{\phi}_{1}{\leq}{\phi}_{2}{\leq}{\ldots}{\leq}{\phi}_{M}<2{\pi}.
\end{equation}
We will now prove that a unitary operator $\tilde{T}$, whose
eigenvalues are of course all non-zero, satisfies Eq.
\eqref{specond1} if and only if the following two conditions are
satisfied:\\

\noindent(i) The distinct eigenvalues of $\tilde{T}$ are the $M$th
roots of unity, for some positive integer $M$ which is a factor of
$D$ and which may be
equal to $D$ itself.\\

\noindent(ii) The distinct eigenvalues of $\tilde{T}$ have equal
degeneracy.\\

We will first prove the necessity of condition (i), following
which we will see that when this condition is satisfied, condition
(ii) is necessary and sufficient for Eq. \eqref{specond1} to hold.

Our proof of the necessity of (i) begins by establishing that, for
each $r$, there is a positive integer $k_{r}{\in}\{1,{\ldots},M\}$
such that
\begin{equation}
\label{eq1} {\lambda}_{r}^{k_{r}}=1.
\end{equation}
To prove this, notice that, from Eq. \eqref{mod1}, we obtain
\begin{equation}
{\lambda}_{r}^{n}=e^{in{\phi}_{r}}
\end{equation}
for every integer $n$.  When $n$ is non-negative, we see from Eq.
\eqref{Rspec} that we must have
${\lambda}_{r}^{n}{\in}\mathrm{spec}(\tilde{T})$. However, the
spectrum of $\tilde{T}$ is finite. In view of this, consider a
particular eigenvalue ${\lambda}_{r}$ and two arbitrary positive
integers $n_{r}$ and $n'_{r}$.  From Eq. \eqref{Rspec}, we see
that
${\lambda}_{r}^{n_{r}},{\lambda}_{r}^{n'_{r}}{\in}\mathrm{spec}(\tilde{T})$.
 The spectrum of $\tilde{T}$ has precisely $M$ distinct
eigenvalues.  So, for fixed $n_{r}$, let us define
$n'_{r}=n_{r}+k_{r}$, where $k_{r}{\in}\{1,{\ldots},M\}$.  There
clearly must be at least one value of $k_{r}$ for which
${\lambda}_{r}^{n'_{r}}={\lambda}_{r}^{n_{r}}$.  When these are
equal, we have $e^{in'_{r}{\phi}_{r}}=e^{in_{r}{\phi}_{r}}$. This
implies that
\begin{equation}
\label{equals1}
e^{i(n'_{r}-n_{r}){\phi}_{r}}=e^{ik_{r}{\phi}_{r}}={\lambda}_{r}^{k_{r}}=1,
\end{equation}
as required.

One important consequence of Eq. \eqref{eq1} is the fact that
\begin{equation}
\label{1spec} 1{\in}\mathrm{spec}(\tilde{T}).
\end{equation}
This follows from Eq. \eqref{Rspec}, which tells us that any
product of eigenvalues of $\tilde{T}$ is also an eigenvalue of
$\tilde{T}$. We simply apply this to Eq. \eqref{eq1}, taking
$R=k_{r}$ and $r_{1},{\ldots},r_{M}=r$.

From this, we see that the ordering of the angles in
\eqref{phiinequ1} implies that ${\phi}_{1}=0$.  We can then update
\eqref{phiinequ1} in the light of \eqref{1spec} to obtain
\begin{equation}
\label{phiinequ2}
0={\phi}_{1}{\leq}{\phi}_{2}{\leq}{\ldots}{\leq}{\phi}_{M}<2{\pi}.
\end{equation}
Another consequence of Eq. \eqref{Rspec} is the fact that, for
each $r{\in}\{1,{\ldots},M\}$,
\begin{equation}
\label{conj}
{\lambda}_{r}^{-1}={\lambda}_{r}^{*}{\in}\mathrm{spec}(\tilde{T}).
\end{equation}
We obtain this in the following way.  We know from Eq.
\eqref{Rspec} and, in the case of $k_{r}=1$, Eq. \eqref{1spec},
that ${\lambda}_{r}^{k_{r}-1}{\in}\mathrm{spec}(\tilde{T})$.
However, it follows from Eq. \eqref{eq1} that
${\lambda}_{r}^{k_{r}-1}={\lambda}_{r}^{-1}$, so we get Eq.
\eqref{conj}.

Let us now use the above observations to prove that the
${\lambda}_{r}$ must be the $M$th roots of unity.  From Eqs.
\eqref{Rspec} and \eqref{conj}, we easily obtain
\begin{equation}
\label{conjprod}
{\lambda}_{r'}{\lambda}^{*}_{r}{\in}\mathrm{spec}(\tilde{T}),
\end{equation}
for all $r,r'{\in}\{1,{\ldots},M\}$.  We now set
$r'=(r\mathrm{mod}M)+1$.  We also write the angular spacings
between neighbouring eigenvalues as
\begin{equation}
{\delta}_{r}=\left\{
\begin{array}{cc}
{\phi}_{r+1}-{\phi}_{r} :&r{\in}\{1,{\ldots},M-1\} \\
2{\pi}+{\phi}_{1}-{\phi}_{M} :& r=M.
\end{array}
\right.
\end{equation}
Combining these definitions and making use of Eq.
\eqref{conjprod}, we obtain
\begin{equation}
\label{deltinspec} e^{i{\delta}_{r}}{\in}\mathrm{spec}(\tilde{T}).
\end{equation}
The mean value of the ${\delta}_{r}$ is $2{\pi}/M$.  Consider now
the smallest of these angular spacings, which we shall denote by
${\delta}_{min}$, which must be nonzero because we are working
with distinct eigenvalues. To fit the $M$ distinct eigenvalues
around the unit circle, we require that
${\delta}_{min}{\leq}2{\pi}/M$. However, we know from Eq.
\eqref{eq1} that $e^{ik{\delta}_{min}}=1$ for some
$k{\in}\{1,{\ldots},M\}$.  It is impossible to satisfy this
requirement for nonzero ${\delta}_{min}$ unless
${\delta}_{min}{\geq}2{\pi}/M$. Combining these two inequalities
gives
\begin{equation}
{\delta}_{min}=2{\pi}/M.
\end{equation}

It is now easy to see that the ${\lambda}_{r}$ must be the $M$th
roots of unity. Given that $e^{i{\delta}_{min}}$ is an eigenvalue
of $\tilde{T}$, which we know to be the case from Eq.
\eqref{deltinspec}, we can apply Eq. \eqref{Rspec} to conclude
that the $e^{ir{\delta}_{min}}$, for all $r{\in}\{1,{\ldots},M\}$,
are also eigenvalues of $\tilde{T}$. These $M$ complex numbers,
which are distinct, are the $M$th roots of unity. Since
$\tilde{T}$ has exactly $M$ distinct eigenvalues, we conclude that
the spectrum of $\tilde{T}$ consists precisely of these $M$th
roots of unity. This completes the proof of the necessity of
condition (i).

Let us now show that when condition (i) is satisfied, condition
(ii) is necessary and sufficient for $\tilde{T}$ to satisfy Eq.
\eqref{specond1}. We will begin by proving its necessity.  The
eigenvalues ${\lambda}_{r}$ of $\tilde{T}$ have been grouped
according to their multiplicity. So, let us denote the degeneracy
of ${\lambda}_{r}$, as an eigenvalue of $\tilde{T}$, by
$d_{r}^{\tilde{T}}$. Combining the fact that the ${\lambda}_{r}$
are the $M$th roots of unity for some integer factor $M$ of $D$
with the phase ordering in Eq. \eqref{phiinequ2}, we see that the
distinct eigenvalues of $\tilde{T}$ are given by

\begin{equation}
\label{evals} {\lambda}_{r}=e^{\frac{2{\pi}i(r-1)}{M}}.
\end{equation}
Furthermore, must have
\begin{equation}
\label{dsum1} \sum_{r=1}^{M}d_{r}^{\tilde{T}}=D.
\end{equation}
Of course, the ${\lambda}_{r}$ are  also the eigenvalues of
$\tilde{T}{\otimes}\tilde{T}$.  However, they will have different
degeneracies.  So, let us denote by
$d_{r}^{\tilde{T}{\otimes}\tilde{T}}$ the degeneracy of
${\lambda}_{r}$ as an eigenvalue of $\tilde{T}{\otimes}\tilde{T}$.
For these degeneracies, we have
\begin{equation}
\label{dsum2}
\sum_{r=1}^{M}d_{r}^{\tilde{T}{\otimes}\tilde{T}}=D^{2}.
\end{equation}
As a consequence of Eq. \eqref{specond1}, we see that
\begin{equation}
\label{ddeq}
d_{r}^{\tilde{T}{\otimes}\tilde{T}}=Dd_{r}^{\tilde{T}}.
\end{equation}
Making use of Eq. \eqref{evals}, we find that the
$d_{r}^{\tilde{T}{\otimes}\tilde{T}}$ can be explicitly expressed
in terms of the $d_{r}^{\tilde{T}}$ in the following way: define
\begin{equation}
\label{Gdef} G_{rss'}=\left\{
\begin{array}{cc}
1 :& (s+s'-r)\mathrm{mod}M=1 \\ 0 :& (s+s'-r)\mathrm{mod}M{\neq}1.
\end{array}
\right.
\end{equation}
where $s,s'{\in}\{1,{\ldots},M\}$.  After some algebra, we find
that we may write
\begin{equation}
\label{Geq1}
d_{r}^{\tilde{T}{\otimes}\tilde{T}}=\sum_{s,s'=1}^{M}G_{rss'}d_{s}^{\tilde{T}}d_{s'}^{\tilde{T}}.
\end{equation}
Combining Eqs. \eqref{ddeq} and \eqref{Geq1}, we see that the
degeneracies $d_{r}^{\tilde{T}}$ must satisfy
\begin{equation}
\label{Geq2}
\sum_{s,s'=1}^{M}G_{rss'}d_{s}^{\tilde{T}}d_{s'}^{\tilde{T}}=Dd_{r}^{\tilde{T}}.
\end{equation}
This is a necessary and sufficient condition for the
${\lambda}_{r}$ to satisfy Eq. \eqref{specond1}.  It is evident
from this expression that, for each $r$, the left hand side is a
quadratic form.  For example, for $r=1$, we have
\begin{equation}
\label{matrix} \left(\begin{array}{c} d_{1}^{\tilde{T}} {\cdots}
 d_{M}^{\tilde{T}}\end{array}\right)
    \left(\begin{array}{cccc} 1 & 0 & {\ldots} & 0 \\
                              0 & 0 & {\ldots} & 1 \\
                              {\vdots} & {\vdots} & \reflectbox{$\ddots$} & {\vdots} \\
                              0 & 1 & {\ldots} & 0
                              \end{array}\right)
    \left(\begin{array}{c} d_{1}^{\tilde{T}}\\ \vdots \\ d_{M}^{\tilde{T}}
    \end{array}\right)=Dd_{1}^{\tilde{T}}.
\end{equation}
The corresponding quadratic forms for $r=2,{\ldots},M$  are
obtained from Eq. \eqref{matrix} by cyclically shifting the
elements of each column in this matrix down by $r-1$ places. Let
us define
\begin{equation}
\label{sigmars} {\sigma}(r,s)=(r-s)\mathrm{mod}M+1.
\end{equation}
Using this and Eq. \eqref{Gdef}, one can readily verify that
\begin{equation}
\label{Geq3}
\sum_{s=1}^{M}G_{rss'}d_{s}^{\tilde{T}}=d_{{\sigma}(r,s')}^{\tilde{T}},
\end{equation}
from which we obtain
\begin{equation}
\sum_{s,s'=1}^{M}G_{rss'}d_{s}^{\tilde{T}}{\delta}_{s'1}=d_{r}^{\tilde{T}}.
\end{equation}
Here, ${\delta}_{s'1}$  is the usual Kronecker delta.  Combining
this equation with Eq. \eqref{Geq2}, we get
\begin{equation}
\sum_{s,s'=1}^{M}G_{rss'}d_{s}^{\tilde{T}}(d_{s'}^{\tilde{T}}-D{\delta}_{s'1})=0.
\end{equation}
Making use of Eq. \eqref{Geq3}, we find that this equation leads
to
\begin{equation}
\label{deq}
\sum_{s'=1}^{M}d_{{\sigma}(r,s')}^{\tilde{T}}d_{s'}^{\tilde{T}}=d_{r}^{\tilde{T}}\sum_{s'=1}^{M}d_{s'}^{\tilde{T}}.
\end{equation}
We will now use this expression to show that the degeneracies
$d_{r}^{\tilde{T}}$ must all be equal to $D/M$.  Notice, from Eq.
\eqref{dsum1}, that $D/M$ is the average of the
$d_{s'}^{\tilde{T}}$. They must all be equal if the maximum
degeneracy is equal to this average degeneracy.  Let $r_{max}$ be
a value of $r$ such that $d_{r_{max}}^{\tilde{T}}$ is the maximum
degeneracy. As a consequence of the positivity of the
$d_{r}^{\tilde{T}}$, the following inequality must be satisfied
\begin{equation}
\sum_{s'=1}^{M}d_{{\sigma}(r_{max},s')}^{\tilde{T}}d_{s'}^{\tilde{T}}{\leq}d_{r_{max}}^{\tilde{T}}\sum_{s'=1}^{M}d_{s'}^{\tilde{T}}.
\end{equation}
with the equality holding only if
$d_{{\sigma}(r_{max},s')}^{\tilde{T}}=d_{r_{max}}^{\tilde{T}}$ for
all $s'$. Now, for any fixed $r$, ${\sigma}(r,s')$ merely permutes
the integers $s'{\in}\{1,{\ldots},M\}$, so that all degeneracies
must, from Eq. \eqref{deq}, be equal to the maximum degeneracy.
This completes the proof of necessity.

Let us finally prove that when the distinct eigenvalues of
$\tilde{T}$ are the $M$th roots of unity, it is also sufficient
that they have equal degeneracies $d_{r}^{\tilde{T}}=D/M$ to
satisfy Eq. \eqref{specond1}. This is simple to show.  For
${\lambda}_{r}$ given by Eq. \eqref{evals}, Eq. \eqref{Geq2} is
equivalent to the spectral copying condition in Eq.
\eqref{specond1}. When $d_{r}^{\tilde{T}}=D/M$, Eq. \eqref{Geq2}
is equivalent to
\begin{equation}
\label{suff} \sum_{s,s'=1}^{M}G_{rss'}=M.
\end{equation}
To show that this equation is satisfied, we note that, when the
$d_{r}^{\tilde{T}}$ are all equal, then Eq. \eqref{Geq3} gives
\begin{equation}
\sum_{s=1}^{M}G_{rss'}=1.
\end{equation}
Summing this expression over the index $s'$ and making use of Eq.
\eqref{Gdef}  leads to Eq. \eqref{suff}, completing the proof of
sufficiency.

Let us take the opportunity here to discuss the above results, in
their physical context, prior to exploring some of their
consequences. For two orthogonal, maximally entangled bipartite
states $|{\psi}_{1}{\rangle}$ and $|{\psi}_{2}{\rangle}$, having
$D$ dimensional subsystems, to be locally copyable with a $D$
dimensional maximally entangled blank state, it is necessary and
sufficient that the eigenvalues of the associated unitary operator
$\tilde{T}$, defined through Eqs. \eqref{tjj} and
\eqref{tildetdef} are, for some integer factor $M$ of $D$, the
$M$th roots of unity and that these eigenvalues are equally
degenerate.

We defined the operator $\tilde{T}$ in Eq. \eqref{tildetdef} in
terms of the operator $T$ which contains all of the information
about the relationship between  $|{\psi}_{1}{\rangle}$ and
$|{\psi}_{2}{\rangle}$.  This definition amounted to the removal
of the phase factor $e^{i{\Delta}{\theta}}$ in Eq.
\eqref{tildetdef}.  This factor was removed in order to simplify
the above proofs of the LOCC copying conditions.  However, for a
particular pair of states, it is $T$, rather that $\tilde{T}$,
that arises naturally. As such, it is important to formulate these
LOCC copying conditions in terms of the spectrum of the $T$
operator also. This is easily done.   The incorporation of this
arbitrary phase factor is equivalent to an arbitrary rotation of
the spectrum in the complex plane.  So, LOCC copying of
$|{\psi}_{1}{\rangle}$ and $|{\psi}_{2}{\rangle}$ is possible if
and only if the eigenvalues of $T$ are, up to an overall rotation,
equally degenerate $M$th roots of unity for some integer factor
$M$ of $D$.  In other words, they must have equal angular spacing
and be equally degenerate.

Clearly, for any particular pair of orthogonal maximally entangled
states $|{\psi}_{1}{\rangle},|{\psi}_{1}{\rangle}$ and a
particular maximally entangled blank state $|b{\rangle}$ for which
the LOCC copying operation is possible, it is important to have an
explicit prescription for carrying out this procedure.  This
amounts to knowing two suitable local unitary operators $A$ and
$B$ for which Eq. \eqref{trans3} is satisfied.  From the results
we have obtained here and in the preceding section, it is possible
to obtain specific operators which carry out the required task.

Our starting point is the three states involved in the copying
procedure, and also the arbitrary reference maximally entangled
state $|{\psi}_{max}{\rangle}$.  These are presumably known. From
these, we deduce the operator $T$ using Eq. \eqref{indep} and the
fact that $T=T_{12}$. The operator $\tilde{T}$ is obtained using
Eq. \eqref{tildetdef} and by setting $-{\Delta}{\theta}$ equal to
the smallest among the arguments of the eigenvalues of $T$.  From
Eq. \eqref{ttrans} and the unitarity of $A$, we see that we may
write
\begin{eqnarray}
\tilde{T}{\otimes}{\id}&=&\sum_{r=1}^{M}{\lambda}_{r}P_{r}, \\
\tilde{T}{\otimes}\tilde{T}&=&\sum_{r=1}^{M}{\lambda}_{r}Q_{r}.
\end{eqnarray}
Here, $P_{r}$ and $Q_{r}$ are the projectors onto the eigenspaces
of ${\lambda}_{r}$, which is an $M$th root of unity given by Eq.
\eqref{evals}, as an eigenvalue of $\tilde{T}{\otimes}{\id}$ and
$\tilde{T}{\otimes}\tilde{T}$ respectively.  Let us denote these
eigenspaces by ${\cal H}_{r}^{\tilde{T}{\otimes}{\idsmall}}$ and
${\cal H}_{r}^{\tilde{T}{\otimes}\tilde{T}}$
 These spaces have dimension $Dd^{\tilde{T}}_{r}$.
Using these notions, we can obtain a unitary operator $A$ that
satisfies Eq. \eqref{ttrans} in the following way.  Let
$\{|{\xi}_{rl}{\rangle}\}$ and $\{|{\eta}_{rl}{\rangle}\}$ be
orthonormal bases for ${\cal H}_{r}^{T{\otimes}{\idsmall}}$ and
${\cal H}_{r}^{T{\otimes}T}$ respectively.  We clearly have
$l{\in}\{1,{\ldots},Dd^{\tilde{T}}_{r}\}$.  Now consider the
unitary operator
\begin{equation}
A=\sum_{r=1}^{M}\sum_{l=1}^{Dd^{\tilde{T}}_{r}}|{\xi}_{rl}{\rangle}{\langle}{\eta}_{rl}|.
\end{equation}
One can easily show that $AP_{r}A^{\dagger}=Q_{r}$, which implies
that $A$ satisfies Eq. \eqref{ttrans} as required.

We must now find a suitable operator $B$.  To do so, we are
required to know the operator $U_{b}$.  This can be deduced from
Eq. \eqref{bdef} to be
\begin{equation}
U_{b}=D{\times}\mathrm{PT}(|b{\rangle}{\langle}{\psi}_{max}|).
\end{equation}
If we now combine Eqs. \eqref{cj} and \eqref{bis}, we find that
$B$ is given by
\begin{equation}
B=e^{i{\theta}_{j}}(U_{j}{\otimes}U_{b})^{T}A^{T}(U^{\dagger}_{j}{\otimes}U^{\dagger}_{b})^{T},
\end{equation}
for either $j=1,2$ and where $T$ again denotes the transpose in
the $|X_{\mu}{\rangle}$.  We may neglect the phase factor here
entirely as it has no affect on the physical nature of the
transformation.

We shall now explore some of the consequences of the local copying
condition in Eq. \eqref{specond1}, paying particular regard to the
relationship between orthogonality and local copyability of two
maximally entangled states with a maximally entangled blank state.

\subsection{Consequences}

 Having established the LOCC copying condition for a
pair of orthogonal, maximally entangled, bipartite, pure states
with a maximally entangled blank state, it is natural to enquire
as to when this condition is satisfied. We shall find that the
dimensionality $D$ of the single particle Hilbert space ${\cal H}$
plays a prominent role here.

We will show that for $D=2,3$, every pair of orthogonal, maximally
entangled, bipartite, pure states can be locally copied with a
maximally entangled blank state. However, we will then show that
for every $D$ which not prime, one can construct pairs of such
states for which this is impossible.

The proof for $D=2$ is a simple matter.  From Eqs. \eqref{overlap}
and \eqref{tildetdef}, we know that the condition of orthogonality
is $\mathrm{Tr}(T)=\mathrm{Tr}(\tilde{T})=0$. For $D=2$,
$\tilde{T}$ has just two, non-degenerate eigenvalues, implying
that $\tilde{T}$ having zero trace is equivalent to these summing
to zero. Writing these two eigenvalues as $e^{i{\phi}_{1}}$ and
$e^{i{\phi}_{2}}$, where we take $0={\phi}_{1},{\phi}_{2}<2{\pi}$
as in \eqref{phiinequ2}, it is easily shown that this
orthogonality condition can only be satisfied if
${\phi}_{2}={\pi}$.  When this is so, they are the 2nd roots of
unity. So, for $D=2$, any pair of orthogonal, maximally entangled
states can be locally copied. This finding is in accord with the
results of Ghosh et al \cite{Ghosh} who showed that with 1 ebit of
entanglement in the blank state, it is possible to copy, by LOCC,
any pair of Bell states.

Let us now consider the case of $D=3$.  Here, the $\tilde{T}$
operator has 3 eigenvalues, $e^{i{\phi}_{1}}, e^{i{\phi}_{2}}$ and
$e^{i{\phi}_{3}}$.  Again we take the phase ordering
$0={\phi}_{1}{\leq}{\phi}_{2}{\leq}{\phi}_{3}<2{\pi}$.  If the
states are orthogonal, then
\begin{equation}
1+e^{i{\phi}_{2}}+e^{i{\phi}_{3}}=0.
\end{equation}
Clearly, this is equivalent to
$e^{i{\phi}_{2}}+e^{i{\phi}_{3}}=-1$. Separating the real and
imaginary parts of this equation gives
\begin{eqnarray}
\label{re}
{\cos}({\phi}_{2})+{\cos}({\phi}_{3})&=&-1, \\
\label{im} {\sin}({\phi}_{2})+{\sin}({\phi}_{3})&=&0.
\end{eqnarray}
From Eq. \eqref{im} we see that
${\sin}^{2}({\phi}_{2})={\sin}^{2}({\phi}_{3})$, which in turn
gives ${\cos}^{2}({\phi}_{2})={\cos}^{2}({\phi}_{3})$ and so
${\cos}({\phi}_{2})={\pm}{\cos}({\phi}_{3})$.  It is easily seen
that we cannot have the minus sign here, since this would
contradict Eq. \eqref{re}. We therefore obtain
\begin{equation}
{\cos}({\phi}_{2})={\cos}({\phi}_{3}).
\end{equation}
Substituting this into Eq. \eqref{re} gives
\begin{equation}
\label{coseqn} {\cos}({\phi}_{2})={\cos}({\phi}_{3})=-\frac{1}{2},
\end{equation}
which implies that ${\phi}_{2}$ and ${\phi}_{3}$ must individually
be equal to either $2{\pi}/3$ or $4{\pi}/3$.  It follows from Eq.
\eqref{im} that these two angles must be different, because Eq.
\eqref{coseqn} implies that the sines of these two possible angles
are nonzero. Combining this with the fact that
${\phi}_{3}{\geq}{\phi}_{2}$, we conclude that
${\phi}_{2}=2{\pi}/3$ and ${\phi}_{3}=4{\pi}/3$.  The three
eigenvalues are then the non-degenerate 3rd roots of unity and Eq.
\eqref{specond1} is satisfied as desired. It follows that the two
states are locally copyable with a maximally entangled blank
state.
\begin{figure}
\begin{center}
\epsfxsize6cm \centerline{\epsfbox{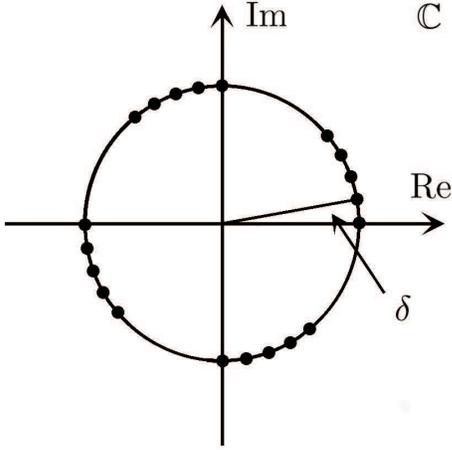}}
\end{center}
\caption{Illustration of the fact that, if $D$ is not prime, then
one can construct a traceless unitary operator whose eigenvalues
are not equally spaced.  The eigenvalues of $\tilde{T}$, which are
represented as points on the unit circle in the complex plane,
have minimum angular separation ${\delta}$. In this example, we
have taken $D_{1}=4$ and $D_{2}=5$.  Here, the tracelessness
condition may be seen to follow from the fact that opposite
eigenvalues cancel each other out and so they all sum to zero.}
 \label{figure2}
\end{figure}

The above analysis shows that for $D=2,3$, any pair of orthogonal,
maximally entangled states can be copied using the same LOCC
operation and a maximally entangled blank state. However, as we
shall now see, this does not hold for arbitrary $D$.  In fact, we
will now demonstrate that for any $D$ which is not prime, one can
construct pairs of orthogonal, maximally entangled states for this
is impossible.

If $D$ is not prime, then, by definition, there exist positive
integers $D_{1},D_{2}{\geq}2$ such that
\begin{equation}
D=D_{1}D_{2}.
\end{equation}
Consider now the $D_{1}$th roots of unity
$e^{\frac{2{\pi}i(j-1)}{D_{1}}}$, where
$j{\in}\{1,{\ldots},D_{1}\}$. The angular spacing between these
complex numbers is $2{\pi}/D_{1}$. Consider now some small angular
interval ${\delta}$ and an operator $T$ with the following set of
distinct eigenvalues
\begin{equation}
\label{ljk}
{\lambda}_{jj'}=e^{\frac{2{\pi}i(j-1)}{D_{1}}}e^{i(j'-1){\delta}},
\end{equation}
where $j'{\in}\{1,{\ldots},D_{2}\}$.  It should be noted at this
point that every unitary operator $T$ on ${\cal H}$ corresponds to
a set of pairs of maximally entangled bipartite states
$|{\psi}_{1}{\rangle}$ and $|{\psi}_{2}{\rangle}$. Indeed, for
arbitrary, fixed $T$ and $|{\psi}_{2}{\rangle}$, we can see from
Eqs. \eqref{psiju} and \eqref{tjj} that $|{\psi}_{1}{\rangle}$ is
obtained using
\begin{equation}
|{\psi}_{1}{\rangle}=(T{\otimes}{\id})|{\psi}_{2}{\rangle}.
\end{equation}

A set of eigenvalues of the form given in Eq. \eqref{ljk} is
depicted in Figure (2), with $D_{1}=4$ and $D_{2}=5$. We can
easily choose ${\delta}$ in such a way that these will not be
equally spaced. We may simply take any ${\delta}<2{\pi}/D$ to
achieve this. However, any unitary operator $T$ whose eigenvalues
are the ${\lambda}_{jj'}$, with these being non-degenerate, can be
seen to be traceless. We have
\begin{equation}
\mathrm{Tr}(T)=e^{-i\left(\frac{2{\pi}}{D_{1}}+{\delta}\right)}\sum_{j=1}^{D_{1}}e^{\frac{2{\pi}ij}{D_{1}}}\sum_{j'=1}^{D_{2}}e^{ij'{\delta}}=0
\end{equation}
because the first sum vanishes.  So, the corresponding states are
orthogonal. However, the fact that the eigenvalues are not equally
spaced implies that the LOCC copying procedure is impossible.

So, we have seen that for $D=2,3$, any pair of orthogonal,
maximally entangled, bipartite, pure states can be locally copied
with a maximally entangled blank state. However, this is not
generally the case when $D$ is not prime. As a consequence of this
finding, a natural question to ask is: for a fixed value of $D$,
is a necessary and sufficient condition for LOCC copying of every
pair of orthogonal, maximally entangled, bipartite, pure states,
with a maximally entangled blank state, the primality of $D$? We
have been unable to determine whether or not this is so.\\

\section{Discussion}
\label{sec5}
\renewcommand{\theequation}{5.\arabic{equation}}
\setcounter{equation}{0}

In this paper, we have addressed the problem of LOCC copying of
entangled states with an entangled blank state.  We were concerned
mainly with the situation where one of the states to be copied is
maximally entangled.  When this is the case, we must have at least
one additional maximally entangled state, and this may be taken to
be the blank state.  When none of the states to be copied are
maximally entangled, it is possible that that the most efficient
use of entanglement occurs when the blank state is incomparable
with the states to be copied.  We illustrated this in section
\ref{sec2}. This is an application of the well known phenomenon of
entanglement catalysis.  There is much work still to be done on
entanglement catalysis before we can have a full understanding of
the process of catalytic copying.

Fortunately, when one of the states to be copied is maximally
entangled, this issue does not arise.  In section \ref{sec3}, we
derived a necessary and sufficient condition for LOCC copying a
set of $N$ states including a maximally entangled state and with a
maximally entangled blank state.  This condition is, in general,
difficult to solve for arbitrary $N$ and subsystem dimension $D$.
However, we were able to make some interesting general
observations about the sets of states that can be copied and the
associated copying transformations. Firstly, if one of the states
to be copied is maximally entangled, then they must all be
maximally entangled. Secondly, without loss of generality, the
copying transformation may be taken to consist of just two unitary
operations, with one being implemented in each laboratory.

For $N=2$, this condition could be solved exactly for all $D$.  We
found that it relates to the eigenvalues of a certain unitary
operator associated with the pair of states to be copied. These
eigenvalues must, up to a phase, be the $M$th roots of unity, for
some factor $M$ of $D$, and they must be equally degenerate.
Having this information enabled us to show that for $D=2,3$, any
pair of maximally entangled, orthogonal states can be copied by
LOCC with a maximally entangled blank state.  However, we were
also able to show that for every $D$ which is not prime, there
exist pairs of such states for which this is not possible.

\section*{Acknowledgements}

FA was supported by a University of Hertfordshire Postgraduate
Studentship.  AC was supported by a University of Hertfordshire
Postdoctoral Research Fellowship. MBP was supported by a Royal
Society Leverhulme Trust Senior Research Fellowship, the EPSRC
QIP-IRC and the EU Thematic Network QUPRODIS.

\end{document}